# ENHANCING CONGESTION CONTROL TO ADDRESS LINK FAILURE LOSS OVER MOBILE AD-HOC NETWORK

Mohammad Amin Kheirandish Fard[1]
Sasan Karamizadeh[1], Mohammad Aflaki[2]

[1]Department of Computer system and Communication, Universiti Teknologi Malaysia
[2] Faculty of engineering, Multi Media University, Cyberjaya, Selangor, Malaysia
`Kheirandish.amin@gmail.com, ksasan2@live.utm.my,`
`mohammad.aflaki06@mmu.edu.my`

## ABSTRACT

*Standard congestion control cannot detect link failure losses which occur due to mobility and power scarcity in multi-hop Ad-Hoc network (MANET). Moreover, successive executions of Back-off algorithm deficiently grow Retransmission Timeout (RTO) exponentially for new route. The importance of detecting and responding link failure losses is to prevent sender from remaining idle unnecessarily and manage number of packet retransmission overhead. In contrast to Cross-layer approaches which require feedback information from lower layers, this paper operates purely in Transport layer. This paper explores an end-to-end threshold-based algorithm which enhances congestion control to address link failure loss in MANET. It consists of two phases. First, threshold-based loss classification algorithm distinguishes losses due to link failure by estimating queue usage based on Relative One-way Trip Time (ROTT). Second phase adjusts RTO for new route by comparing capabilities of new route to the broken route using available information in Transport layer such as ROTT and number of hops.*

## 1. Introduction

Mobile Ad-Hoc network is a collection of mobile nodes which form a self-organizing and self-configurable network that has no central administrator. In contrast to infrastructural wireless network which use base station to manage nodes in its area, MANET does not require any fixed infrastructure. Nodes within same transmission range can communicate directly while those are not can use other nodes as relays to send packet. Inexpensive deployments of MANET due to absence of fixed infrastructure as well as mobility feature for all nodes have considered MANET as a subject of researches. However, MANET consists of unstable wireless communication links in compare to the wired network that produces new kind of losses.

Three model of loss exist in network connection. Congestion is identified as major causes of packet loss in wired network. However, wireless characteristics such as interference of radio signal, radio channel contention and low bandwidth can lead wireless link unreliable. Link failure mostly occurs when mobile node which forms a route launches to move out of its neighbourhood's transmission range. In addition, battery depletion can make link breakage. Thus, in addition to congestion, link failure and wireless channel error have significant contribution in generating loss in MANET.

Congestion control is the most controversial parts of TCP which degrades performance when encounters non-congestion loss in MANET. Congestion control assumes all loss induced by congestion. For example, link breakage lasts greater than RTO is misinterpreted as congestion loss. Thus regardless of kind of loss, it decreases sending rate to alleviate congestion and grows retransmission timeout exponentially to wait more for receiving acknowledgment. It is plausible in wired network since non-congestion loss occurs rarely and also some application can tolerate some degrees of error. However, this unnecessary throughput drop which waste available resources such as bandwidth arises in MANET.

Link failure needs TCP to explore how much new route is congested in compare to the broken one. Traffic characteristics can affect queuing delay and processing delay of intermediate nodes that consequently influences Round Trip Time (RTT). If discovered route suffers heavier traffic than old one, retransmission timer must wait more to receive acknowledgment and RTO should be increased. Otherwise, when new route is approximately non-congested, data packets and acknowledgment transferred quicker than old route. Thus sender must wait less than before to receive acknowledgment and RTO should decrease.

Figure 1 shows, how sender gradually alters retransmission timeout when route changes. For sake of simplicity, required time for finding new route is ignored. Sender launches to transmit packet in route1. After 'a' seconds, route failure happens. Then, retransmission timer expires, packet loss detected and sender retransmits last unacknowledged packet. When sender receives acknowledgment of retransmitted packet, it implies route2 is discovered for transmission. During time (a, b) which is called adaption period, sender closes RTO to its actual value after receiving some RTTs from route2. At 'b', sender finds stable RTO for route2.

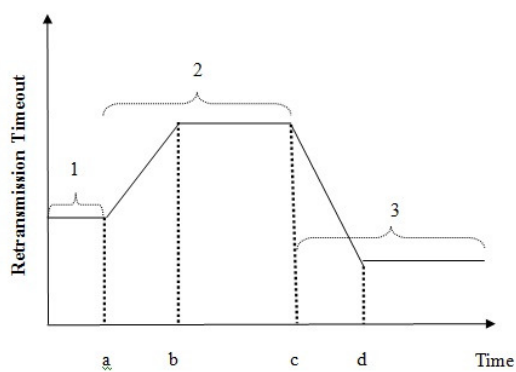

Figure 1.Retransmission timeout varies over link breakage

TCP gradually adapts RTO to its actual value after receiving some RTTs from reconstructed route in adaptation period. If any packet loss happens in adaption period (a, b), sender uses short RTO of route1 while transmitting data in route2 whose needed RTO is longer. Therefore retransmission timer expires frequently and leads to unnecessary retransmission overhead which boosts traffic. Second serious problem emerges when packet loss during (c, d) forces sender to use long RTO of route2 while route3 requires shorter RTO. Any packet loss needs sender reacts quickly while it still uses long RTO of route2. Thus, sender will be idle unnecessarily.

Thus, efficient loss differentiation algorithm (LDA) helps TCP to identify loss causes and consequently triggers appropriate loss recovery algorithm (LRA) over each kind of losses.

## 2. Literature Review

Sender should recognize state of MANET and wireless link to act accordingly. For example, specifying available buffer of intermediate nodes that assess congestion can greatly influence recovery operations. Measuring remained energy of nodes can assists sender to change route before link breakage. Calculating distance between nodes based on signal strength can help sender to predicate future link failure and switch into another route before breakage. All these information can be either measured explicitly with support from intermediate nodes or estimated implicitly from information in received acknowledgment.

In First mechanism, TCP sender entirely does the job and estimate MANET situation implicitly without any support from intermediate node. It does not create processing overhead at intermediate routers. The main drawback of it, is the lack of detailed information about state of wireless link at the sender [15].For example, Fixed RTO interprets two successive timeout as route failure. Then retransmit unacknowledged packet while it keeps value of RTO unchanged [8]. However two successive timeout can be sign of congestion in congested MANET and is highly based on existing traffic pattern. That`s why it is not precise enough.

In feedback (cross layers) approaches, sender get detailed information from network state by collaborating between TCP layers of intermediate nodes. For example, since congestion control is not aware from losses due to wireless medium contention over 802 MAC protocol, it must collaborate with MAC layer to address these losses [6] [17]. Although Feedback methods are more precise than end-to-end approach [1], modifications in intermediate nodes make implementation complicated for WAN. Moreover, extra overhead produced due to transmission notification packet. In addition, it reduces flexibility [16]. For example, TCP Muzha forces Intermediate nodes to fill special field in acknowledgment header to clarify sender how many empty rooms are available in their buffers [10]. TCP-F [7] and TCP-BUS [9] are feedback approaches which pursue the same mechanism. When intermediate node detects link breakage, route notification message informs source to stop sending further packets and freeze state variable such as RTO. When route rebuilt, route reconstruction notification packet informs source to resume transmission with old RTO.

WestwoodVT is an end-to-end approach which classifies packet loss by estimating existing data packet in buffer of intermediate nodes. It is too resemble to TCP-Veno [13].Actually both inherit policy of TCP Vegas to differentiate causes of packet loss [18]. After received acknowledgment, They measures the difference between expected rate and actual rate and assign it to $\Delta$ which is indication of amount of buffer in queue of middle nodes. Interpreting causes of loss is done based on two predefined threshold $\alpha$ and $\beta$ and available buffer of intermediate nodes as $\Delta$. If it becomes smaller than $\alpha$, buffers of intermediate nodes still can accommodate incoming packets. So WestwoodVT relates any loss due to the wireless error. If $\Delta$ is larger than $\beta$, it shows that buffers are approximately full and any packet loss is due to congestion [18]. If estimated $\Delta$ becomes between two thresholds, decision is postponed to next losses. Main drawback of WestwoodVT that degrades performances (throughput and energy consumption) is revealed when Bit Error Rates (BER) increases [2]. In addition WestwoodVT cannot address link failure. TCP-Feno introduces another challenge on TCP VEGAS proponents (WestwoodVT and TCP-Veno). It claims TCP VEGAS performance degrades in network when nodes use small buffer size [11]. MANET with nodes carrying small buffer size can quickly enter into congestion mode. However, since TCP VEGAS does not contribute maximum buffer size in estimation, it just compares $\Delta$ with threshold $\alpha$ and find out it is less than it. So TCP VEGAS declares loss as non-congestion loss while congestion exist. However TCP-Feno still cannot cope with two first mentioned problems.

Neural network classifier as soft computing solution is an offline end-to-end method for differentiating packet loss when trained with required data set [12]. Trained classifier can be used for unknown situations to predicate type of packet loss. Data set can be extracted from different topologies and random traffic patterns which generate around 5000 packet loss. Each row of data set includes Round trip time, vibration of round trip time, congestion window size and number of repeated acknowledgment as an input and type of packet loss as an output. Required structure includes four inputs nodes, five hidden nodes and three output nodes to produce three different kinds of outputs [12]. Data set is divided into two parts. First group for learning process and second one is for evaluating efficiency. Neural network is trained in offline mode with first group. Then its classification ability is evaluated over second group. Reaching to acceptable error rate makes it appropriate for predicating loss in MANET. Sender transmits packet to destination and save required input variables of recent packet. In case of probable loss, it feeds neural network with saved input variables to determine what kind of loss occurred [12].However, Not only collecting information is overwhelming for implementation, some topologies or traffic patterns might not be covered by data set. In addition, it cannot address link failure problem.

One group of classification approaches compares a classification metric by a threshold to distinguish congestion loss from non-congestion loss. Comparison between threshold and metric play main role in classification process. These mainly focus on packet losses recognized through third duplicate acknowledgements. JTCP is a jitter-based method which calculates average of coming jitter in each round trip time.

$$Jr = \frac{(R_{\,newest} - R_{\,oldest}) - (S_{\,newest} - S_{\,oldest})}{R_{\,newest} - R_{\,oldest}} \qquad (1)$$

S and R represent time of sending and receiving packet respectively and 'newest' and 'oldest' index denote to latest and oldest ack packet. In case of packet loss through third duplicate acknowledgements, if Jr becomes greater than inverse value of congestion window and triple ACKs does not receive in one RTT, congestion has made packet loss. [14]

LDA_RQ is an implicit end-to-end approach which tries to estimate queue usage rate of intermediate nodes. It does not need any support or feedback from middle nodes. Available Information in transport layer are congestion window size (cwnd), round trip time (RTT) [3]. It defines two loss classification formulas, one for beginning and another for rest of transmission. It compares first classification metric with threshold until maximum EROTT exceeds three times greater than minimum EROTT. After this gap appeared, it uses second classification till the end. In addition, special ROTT called congestion ROTT calculated to show border between normal and congested MANET. When TCP recognizes loss through third duplicate acknowledgements, it verifies weather queue usage exceeds 50% or current ROTT becomes greater than congestion ROTT. If either former or latter satisfied, detected loss is due to congestion. Otherwise loss is induced by non-congestion factors [3]. However it cannot detect link failure. In addition, gap between minimum and maximum of EROTT achieved experimentally that definitely vary based on experiment. Moreover, in situation which gap cannot reach to three, queue usage remains less than 30% and Congestion EROTT might not be initialized.

TCP-welcome is an implicit end-to-end scheme which differentiates causes of packet loss based on history of Round Trip Time. Ascending growth of RTT increment is induced by congestion. However, If RTT didn't fluctuate and remained around averaged value, the way packet loss recognized becomes important. Three duplicate acknowledgements are a consequence of wireless channel error while retransmission timeout is due to link failure [2]. However, TCP-WELCOME uses RTT which includes both delays of forward and reverse path while only delay of forward path must be considered. In addition, it offers recovery method based on RTT comparison. TCP-Welcome claimed that RTO adjustment should be done based on the capabilities of discovered route such as length, load and link quality. After link breakage, total delay for new route varies from broken route. Hence, RTT comparison seems to be suitable parameter for tuning RTO.

$$\frac{\text{RTO}_{\text{new}}}{\text{RTO}_{\text{old}}} = \frac{\text{RTT}_{\text{new}}}{\text{RTT}_{\text{old}}} \tag{2}$$

However, RTT is not enough for depicting capabilities of discovered route. In addition, it includes both delay of forward and backward path.

ABRA does not offer method to classify packet losses. However, it uses smoothed Round Trip Time instead of RTT to set RTO after link breakage. When link failure lasts more time than RTO, timeout happens. Standard TCP grows RTO exponentially due to multiple successive back-offs. When route come back, TCP cannot retransmit last unacknowledged packet since it must wait until this long RTO expires. Thus it is serious deficiency since route recovered but TCP remains idle unnecessarily. ABRA claims that new RTO is dependent on the smooth round trip time (SRTT) [5].

$$\frac{\text{RTO}_{\text{new}}}{\text{RTO}_{\text{old}}} = 1 + \frac{\text{Last}_{\text{srtt}} - \text{Min}_{\text{srtt}}}{\text{Max}_{\text{srtt}} - \text{Min}_{\text{srtt}}} \tag{3}$$

## 3. Proposed Method

Existing solutions in loss classification area tried to just differentiate congestion loss from non-congestion loss since it doesn't matter what kind of non-congestion loss (link failure or wireless channel error) occurred. They only want to invoke congestion control over losses due to congestion. In contrast to proposed approaches, this paper studies problem deeply by determining whether link failure loss occurred. Link failure and network portioning which mainly created by factors such as mobility and battery depletion has negative effect on MANET performance. An ideal LDA must classify losses related to link failure from others without imposing additional overhead of notification packets transmitted between mobile nodes [2].

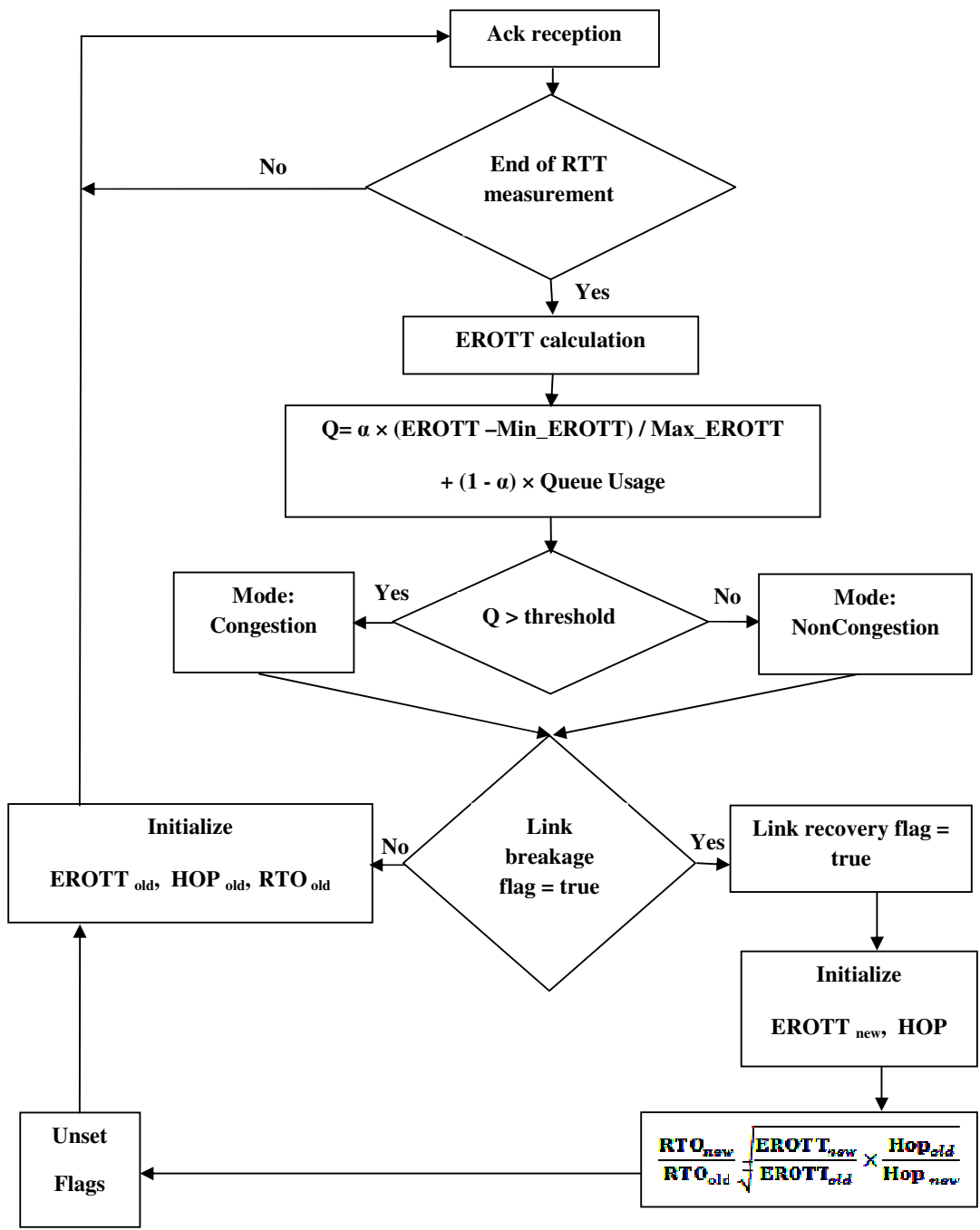

Figure 2. Enhanced Mechanism for Ack Reception

As figure 2 depicted, when enhanced congestion control receives acknowledgment which is not related to end of RTT measurement, nothing is done. Otherwise, ROTT must be calculated and compared with previous maximum and minimum ROTT to reinitialize in case. Then queue usage estimated and compared with threshold. Whenever it exceeds predefined threshold, it is sign of congestion and enters MANET into the congestion mode. Otherwise non-congestion becomes MANET mode. Then it checks link breakage flag. Finding this flag false is an indication of normal MANET that leads in saving current ROTT, RTO and number of HOP related to current route and waiting for next acknowledgment. This cycle repeats until congestion control finds out that link failure flag has been already set. It waits until receives acknowledgment which is an indication of route reconstruction and consequently set link recovery flag. Then it initializes ROTT and number of Hop of new route. At this point, all unknown parameters for adjusting RTO are identified. At the end, TCP unset flags related to link failure to declare normal MANET and save current ROTT, RTO and Hop number.

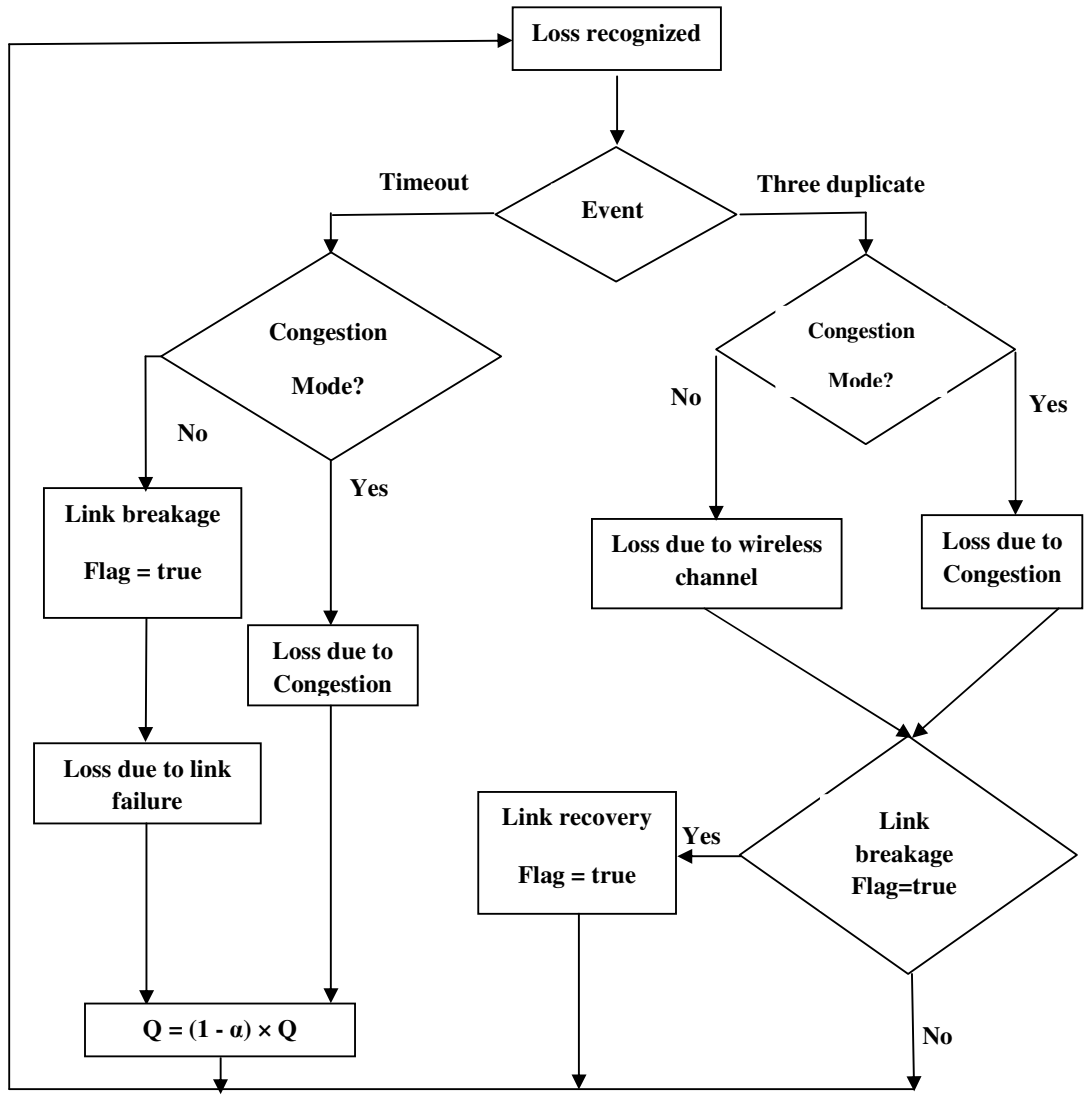

Figure 3. Loss Classification Diagram

In contrast to previous approaches, RTO expiration is addressed in this paper (As shown in figure 3). When timeout occurs, neither data packets nor acknowledgment can reach to the end host that intensifies probability of link breakage. However, receiving duplicate acknowledgments prove the existence of route. Thus in case of Timeout, when probability of congestion rejected, link breakage becomes more likely. At the end of flowchart, since no acknowledgment received in case of Timeout, first part of queue usage equation is zero (shown in figure 2) and only small fraction of recent queue usage considered for current queue usage. Obviously, successive timeout decrease any large ratio to become below than threshold and eventually set link failure flag. As Figure 3 shows, receiving duplicate acknowledgment will set route recovery flag if route failure flag has been already set.

## 3.1 Queue Usage Estimation

Estimating rate of queue usage signifies how much buffers of intermediate nodes are involved. Being less than threshold signifies MANET is not congested and probably, loss is due to wireless factors. Otherwise loss is induced by congestion. The minimum (/maximum) value of EROTT represents emptiness (fullness) of intermediate node`s buffer which. It seems increment or decrement of EROTT becomes a plausible sign for queue usage evaluation.

$$Q = \text{EROTT} \tag{4}$$

Comparing just EROTT to the fixed threshold is not sensible since rebuilt route after link breakage might include more (/less) hop which increases (/decrease) delay and EROTT. Therefore it is divided by MAX_EROTT to depend EROTT based on characteristics of route.

$$Q = \frac{\text{EROTT}}{\text{Max\_EROTT}} \tag{5}$$

However, EROTT and Max_EROTT are almost equal at the beginning and reach to their actual values along with each other. Dividing two equal numbers gives ratio close to one which means queue usage is in maximum while MANET is at starting point. Thus decreasing Min_EROTT from EROTT makes differences between numerator and denominator.

$$Q = \frac{\text{EROTT} - \text{Min\_EROTT}}{\text{Max\_EROTT}} \tag{6}$$

At starting point, Min_EROTT and EROTT are approximately equal. Hence numerator is zero while MAX_EROTT as denominator is greater than zero. Thus usage rate is approximately zero which is correct for starting point. As times passes, Min_EROTT remains stable, EROTT and MAX_EROTT increase to reach to actual value. The queue usage rate gradually increases that signifies MANET is going to be congested. Since in case of Timeout no acknowledgment received, queue usage is zero. Thus, recent queue usage contributed in calculating current queue usage by adding second part in following formula. Therefore, queue usage is no longer zero in case of timeout. In addition, it can alleviate effect of sudden changes of EROTT and close assessment to recent situations.

$$Q = \alpha \times \frac{\text{EROTT} - \text{Min\_EROTT}}{\text{Max\_EROTT}} + (1 - \alpha) \times Q \tag{7}$$

Assigning $\alpha$ in range (0,1) can specify how percentage of recent Q can affect current Q. For estimating Q in case of acknowledgment reception, $\alpha$ is 0.8.Thus only 20% of recent Q used for current estimation. For estimating Q in case of Timeout (As shown at bottom of figure 3) $\alpha$ is 0.1.Thus 90% of recent Q used. Therefore in worst situation (not practical) which MANET experience heavy traffic (Q =1), after six successive Timeout, Q decline to less than 0.5.

## 3.2 RTO Adjustment

Enhanced congestion control reinitializes RTO to adapt it to the discovered route capabilities. Following comparisons show the derivation of final formula step by step. Variables with 'old' and 'new' indexes are related to the broken and re-established route, respectively. RTT increment can originate due to heavy traffic which raises total delay. As a result, TCP should wait more than old route to receive acknowledgment and RTO increases. When discovered route requires larger (/shorter) RTT, retransmission timeout increases (/decreased). (8)

$$\frac{RTO_{new}}{RTO_{old}} = \frac{RTT_{new}}{RTT_{old}} \qquad (8)$$

However, other factor such as number of Hop should be included in RTO adjustment. When sender receives acknowledgment from reconstructed route and calculates RTT, TCP compares it with old RTT of broken route. If new RTT is larger, it signifies that discovered route is more congested than broken route. So, sending rate which is dependent on congestion window size must be decreased to prevent congestion formation (9).

$$\frac{CWND_{new}}{CWND_{old}} = \frac{RTT_{old}}{RTT_{new}} \qquad (9)$$

The number of hops can affect route capabilities since packet accommodation is directly proportional to the number of hops. If discovered route includes more (/less) hops than broken route, congestion window size should be increased (/decreased) (10).

$$\frac{CWND_{new}}{CWND_{old}} = \frac{Hop_{new}}{Hop_{old}} \qquad (10)$$

Equation (9) and (10) indicate number of hops and RTT are inversely proportional to each other. Because, either RTT increment(/decrement) or hop decrement (/increment) forces sender to decrease(/increase) CWND. Hence, instead of RTT in(8), inverse ratio of hop number is replaced (11).

$$\frac{RTO_{new}}{RTO_{old}} = \frac{Hop_{old}}{Hop_{new}} \qquad (11)$$

By combining (8) and (11) and replacing RTT by EROTT, following formula is achieved.

$$\frac{RTO_{new}}{RTO_{old}} = \sqrt{\frac{EROTT_{new}}{ER0TT_{old}} \times \frac{Hop_{old}}{Hop_{new}}} \qquad (12)$$

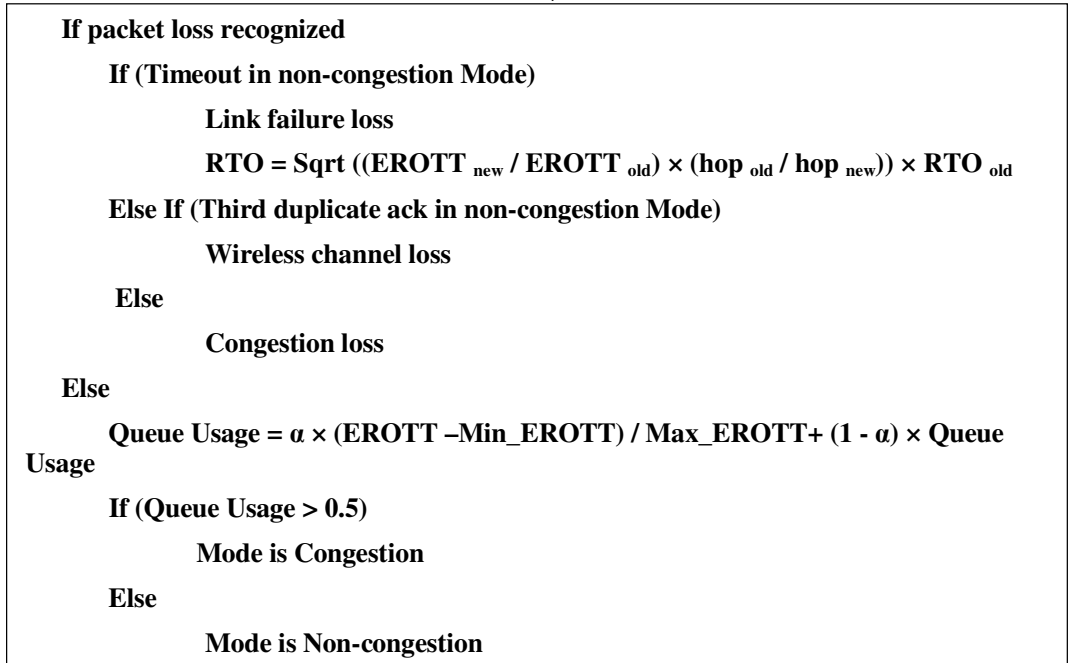

Figure 4. Main Algorithm for Enhancing Congestion Control

## 4. Simulation and Result

Simulation is done by NS2. Radio transmission range is 250 meters, bandwidth of the wireless channel is 2Mbps, MAC layer protocol is 802.11b, and queuing policy is DropTail. Packet size equals to 1Kbytes. FTP connections generate data packets at the first node and continually forward them to reach last node. The packet size is 1000 bytes and the routing protocol used in the simulation is DSDV.

Not only accuracy of classifying loss due to link failure should be satisfactory, but also accuracy of classifying loss due to congestion and wireless channel should be desirable. The accuracy is evaluated by how percentages of losses are distinguished correctly. 6-hop topology of IEEE 802.11 wireless nodes is used to examine Loss classification accuracy of approaches. In first scenario, no link failure loss exists since all nodes are fixed. Wireless error rate is zero. Increasing flow number leads to more congestion loss that should be distinguished from other types of loss. In second scenario, no congestion loss exists since only one flow generates packet. All nodes are fixed. Only wireless error rate changes from low to high level.

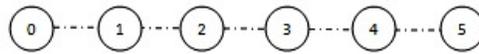

Figure 5. 6-Hop Topology Chain for Loss Classification

In order to evaluate accuracy of link failure classification and verify ability of RTO adjustment, topology shown in figure 6 is used. Node1 which moves to different places causes link failure. Rests of the nodes are fixed. FTP connections which generate data flow from node0 to node5 experience three different routs (0,2,1,3,5),(0,1,5) and (0,1,4,5) respectively.

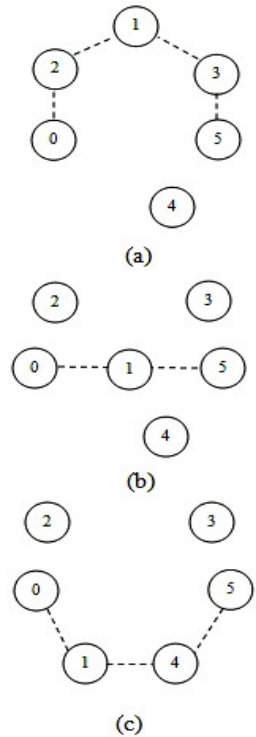

Figure 6. Required Topology for Addressing Link Failure Loss

Since LDA_RQ proposed efficient classification loss among others such as NCPLD, Veno, West, JTCP and RELDS [3] it is chosen for comparison. However, for triggering LDA_RQ, a gap between minimum and maximum of EROTT should reach three while three achieved experimentally and is highly depend on different experiments. Moreover, when gap cannot reach to three, queue usage remains less than 30% and some variables might not be initialized. In addition, since it cannot deal with link failure loss, TCP-welcome added to the experiment.

## 4.1 EROTT vs. RTT

RTT includes delays of forward and backward path. Delay of forward path which carries data packets is related to the input queue while backward path`s delay is related to delay of acknowledgment packets suffer in output queue. As previous study [4] suggested by experiments, EROTT (Estimated Relative One-way Trip Time) can only measure delay of forward path and is more precise than RTT .Since traffic characteristics of output queue vary from input queue, their delays are different and sender cannot halve RTT to estimate EROTT. For example, deploying delayed acknowledgment mechanism which calculates optimum delay windows size for generating acknowledgment at receiver side makes differences between numbers of data packet and acknowledgment. As figure 7 demonstrates, EROTT is always smaller than RTT, but it is not exactly half of it as some points proves it.

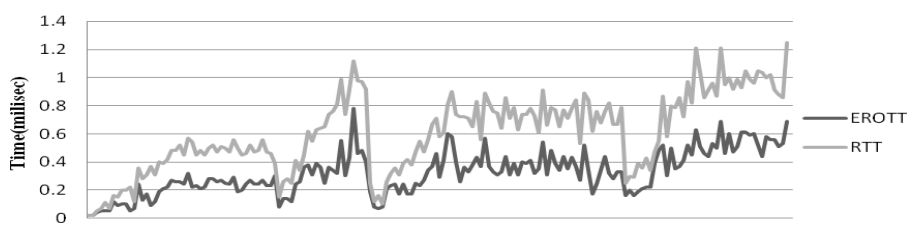

Figure 7. Comparison between RTT and EROTT

## 4.2 Accuracy of Congestion Loss Classification

In first experiment ability of enhanced congestion control in classifying congestion loss is examined. Queue size equals to 50kb. By increasing number of flows, losses due to congestion appear more than before. As figure 8 demonstrates, although classification accuracy of all decreases, enhanced TCP still distinguishes congestion loss better than others.

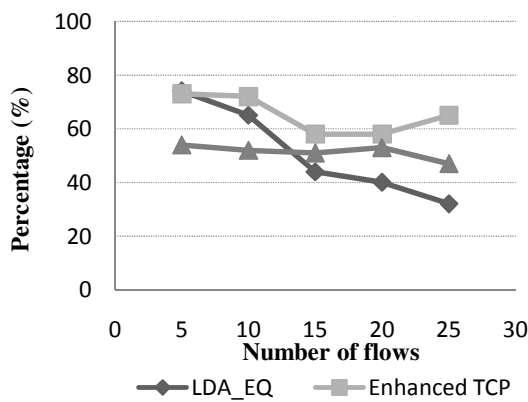

Figure 8. Accuracy of Congestion Loss Classification

Increasing number of flow intensifies probabilities of congestion losses and generates lots of Timeout. Successive timeouts gradually decrease queue usage as equation 13 shows. ($\alpha = 0.1$)

$$Q = (1 - \alpha) \times Q \qquad (13)$$

Thus queue usage mostly remained below than threshold (0.5) and most of the losses are interpreted as non-congestion loss. That's why accuracy of all congestion loss classification decreases by increasing number of flows.

TCP-Welcome only compares new RTT with previous one. Being even a little bit less (/greater) than previous RTT declares MANET in normal (/congested) state no matter MANET has heavy or light traffic. Thus increasing or decreasing number of flow does not affect its accuracy greatly and always fluctuates around average 50%. LDA_RQ launches to operate when gap (=MAX_ROTT/Min_ROTT) exceeds three. Not only this gap varies for different topologies and traffic patterns, but also when this gap cannot reach to three, it prevents LDA_RQ to detect congestion.

### 4.3 Accuracy of Wireless Loss Classification

In second part, wireless channel mainly generate loss during transmission since the only flow in simulation cannot generate congestion losses. Loss rate increases gradually to check whether enhanced congestion control still can achieve best efficiency. Figure 9 shows that enhanced congestion control classifies losses due to wireless channel better than others.

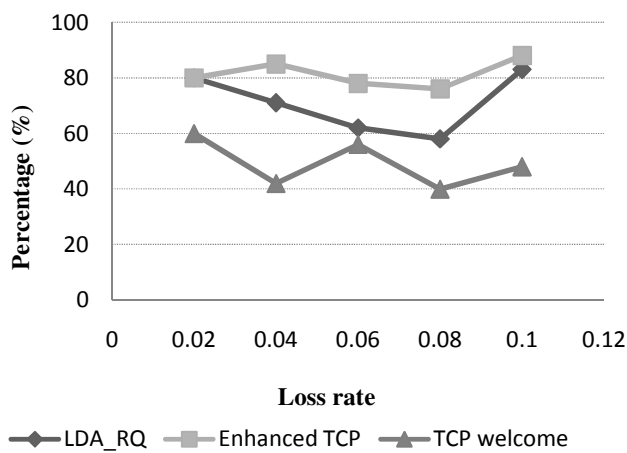

Figure 9. Accuracy Of Wireless Loss Classification

Losses due to wireless channel inserted randomly that do not follow specific pattern. That's why its accuracy is not predictable and fluctuates for various loss rates.

### 4.4 Accuracy of Link Failure Loss Classification

As it is mentioned, main defect of LDA_RQ is its inability to detect link failure. TCP-WELCOME is an approach which addresses link failure based on RTT. It interprets any increment in RTT as congestion while reconstructed route may include more hops which have incremented RTT. Thus, sometimes loss classification process misinterprets losses related to link failure as congestion loss. Since mobility is the main reason of link failure, node1`s speed gradually has increased to investigate how different approaches behave. Figure 10 illustrates enhanced TCP exposes growing accuracy while TCP-WELCOME does not.

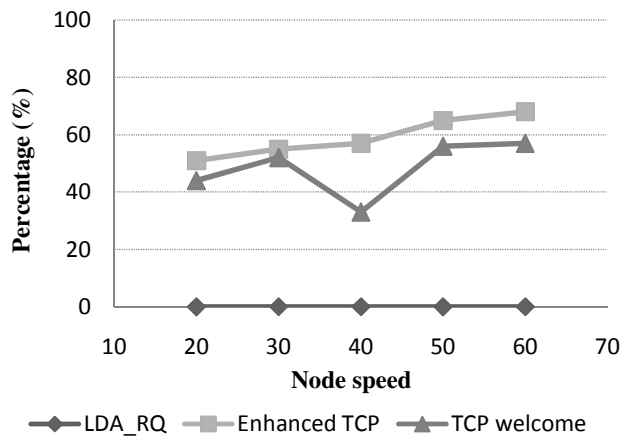

Figure 10. Accuracy of Link Failure Loss Classification

As it is already mentioned, LDA_RQ is not able to detect link failure losses. TCP-welcome always interprets increase in RTT as congestion rather than as a result of switching between routes whose Hop numbers are different. At high speed mobility, link failure emerges quickly while low speed mobility takes much time for node1 to stand out of its neighbor`s transmission range. As a result more number of successive timeout happen at high speed. The more successive timeout happens, the faster MANET enters into non-congestion mode (queue usage< 0.5) and high percentage of future losses are interpreted due to link failure. In addition, in compare to FixedRTO which interpret two successive timeout as link failure and cannot work efficiently in congested MANET, enhanced TCP declare link breakage after five or six (adjustable ; based on α) Timeout.

## 4.5 Throughput

Throughput refers to the number of packets successfully reaches to destination in time unit. Figure 11 illustrates the comparison between different versions of TCP. Measured average throughput under different conditions demonstrates that Enhanced TCP outperforms TCP Reno.

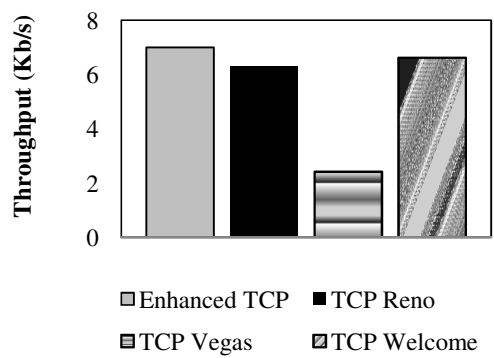

Figure 11. Throughput of Different TCPs

Standard TCP does not interpret successive timeouts and executes back-off algorithm for each timeout occurrence .However specific number of successive timeouts (varies based on topology and traffic pattern) in enhanced TCP extremely decline queue usage to become less than 0.5.

Thus future recognized losses via timeout are interpreted as a result of link failure. Since enhanced TCP is aware of link breakage, it adjusts RTO for rebuilt route while standard TCP resumes using current long RTO for rebuilt route.

In order to evaluate accuracy of loss classification, number of flows increased to raise number of congestion losses. Enhanced TCP should distinguished losses due to link failure from congestion loss and consequently tune RTO. As Figure 12 shows, although throughput of all approaches increases, enhanced TCP still show the best throughput.

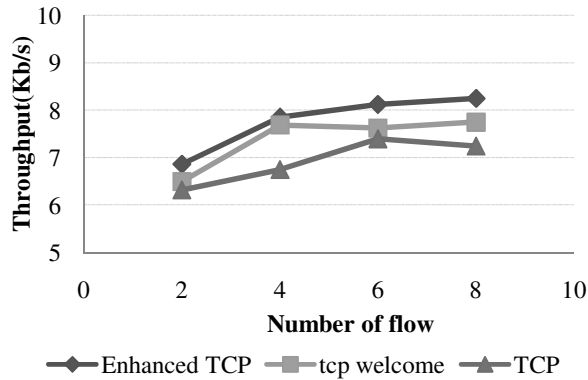

Figure 12. Throughput Comparisons Under Different Flow Number

Obviously by increasing number of flow (definitly limited to certain point), all can better utilize available bandwidth and throughput increase. However TCP cannot detect link failure losses from growing number of congestion loss and operate poorly while Enhanced TCP can.

## 4.5 TCP idle time

This improvement is mainly related to the reduction of long RTO into optimized RTO for discovered route. New term Sum_RTO is introduced which refers to the summation of RTO for all transmitted packet. Having small Sum_RTO with high throughput signifies that respective approach utilizes mentioned idle time efficiently. Figure 13 demonstrates that Sum_RTO for new approach is less than others. TCP Vegas represents the worst throughput among all in figure 11. Thus, its small Sum_RTO due to small number of transferred packets was predictable and it does not signify TCP Vegas act efficiently than others.

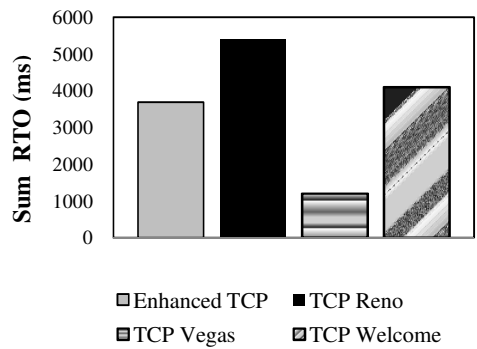

Figure 13. Sum_RTO For Different TCPs

Data packet which is supposed to transmit immediately after link failure contains long RTO in standard TCP. However since enhanced TCP aware of link failure, RTO optimized based on RTT and number of Hop of new route and great portion of it eliminated for transmitting packet while standard TCP resume using long RTO derived from successive execution of back-off algorithm.

As it is mentioned, retransmission overhead exists when rebuilt route requires greater RTO than old one. However, conventional TCP reaches maximum value for RTO after few successive timeouts and remains fixed since it assumes greater RTO practically impossible. Thus simulating a route whose RTO becomes larger than maximum RTO, is not possible. Comparison must be done by approaches which freeze TCP connection in case of link failure and resume transmission with old RTO after receiving notification. However due to their complex implementations, they are not covered in this paper.

## 5. Conclusion

This paper explores an end-to-end threshold-based algorithm which enhances congestion control to address link failure loss in MANET. It consists of two parts. Threshold-based loss classification algorithm uses queue usage to classify network state periodically into congestion or non-congestion mode. Any retransmission timeout in period which MANET is non-congested mode is an indication of link failure loss. In addition, implementation showed that small percentage of three duplicate acknowledgments which emerge immediately after route recovery might be result of route changes. After detecting losses due to link failure, it should adjust RTO for reconstructed route by comparing its capabilities with broken route using available information in transport layer. This enhances congestion control by transmitting packet as soon as route recovered rather than being idle unnecessarily. Simulation results are evaluated in term of three metrics. Loss classification accuracy as a first metric evaluate threshold-based algorithm. Accuracy in classifying loss due to congestion (AC), link failure (AL) and wireless channel (AW) should be enough satisfactory to prevent misinterpreting other losses as link failure loss. Some factors such as mobility, flow number and loss rate are gradually changed to study how AC, AW and AL behave respectively. Since enhanced TCP efficiently uses idle time to transmit packet, throughput and idle time should be evaluated as well. Having high throughput with low idle time signifies improvement. Results showed that enhanced congestion control outperform other approaches.

## REFERENCES

[1] Mi-Young Park, Sang-Hwa Chung "Analyzing Effect of Loss Differentiation Algorithms on Improving TCP Performance" the 12th IEEE International Conference on,(ICACT), Volume: 1 Page(s): 737 – 742, Phoenix Park 2010

[2] Seddik-Ghaleb,A.;Ghamri-Doudane,Y.; Senouci, S.-M. "TCP WELCOME TCP variant for Wireless Environment, Link losses, and Congestion packet loss Models". IEEE Conferences on Communication systems and Networks. Page(s): 1 – 8, Bangalore 2009

[3] Mi-Young Park; Sang-Hwa Chung; Sree kumari."Estimting Rate of Queue Usage to differentiate Cause of Packet Loss in Multi-hop Wireless Networks".Computer Software and Applications,COMPSAC'09.33rd Annual IEEE International Conferences . Volume: 1, Page(s):500–508. Seattle,WA 2009

[4] Vern Paxson, "End-to-End Internet Packet Dynamics," IEEE/ACM Transactions on Networking,Volume: 7, Page(s): 277 – 292. June 1999

[5] Ghanem T.F, Elkiliani, W.S, Hadhoud ,M.M "Improving TCP performance over Mobile Ad Hoc Networks using an adaptive backoff response apparoach" IEEE Conferences on Networking and Media Convergence, ICNM Page(s): 16-21 Cairo 2009.

[6] Kai Chen,Yuan Xue,Nahrstedt K"On Setting TCP Congestion Window Limit in Mobile Ad Hoc Networks" ICC '03.IEEE International Conference .p:1080– 1084, 2003.

[7] K. Chandran, S. Raghunathan, S. Venkatesan and R. Prakash, "A feedback based scheme for improving TCP performance in ad-hoc wireless networks", IEEE JOURNALS Personal Communications Magazine, Volume: 8Page(s): 34–39, Amsterdam 2001

[8] T. Dyer and R. Boppana. "A comparison of TCP performance over three routing protocols for mobile ad hoc networks" In Proceedings of the 2001 ACM International Symposium on Mobile Ad Hoc Networking & Computing (MobiHoc'01, Long Beach, pp.56-66 CA, USA, , 2001

[9] D.Kim, C.K. Toh, Y. Choi, "TCP-BUS: Improving TCPperformance in wireless ad hoc networks", IEEE International Conference on Cominunications Vol 3. Page(s): 1707 – 1713, New Orleans, LA, USA ,2000

[10] Yao-Nan Lien; Ho-Cheng Hsiao (2007). "A New TCP Congestion Control Mechanism over Wireless Ad Hoc Networks by Router-Assisted Approach".27th IEEE International Conference on Distributed Computing Systems Page(s): 84. 2007

[11] Jae-Hyun Hwang,See HwanYoo,Chuck Yoo.(2009). "TCP Feno: Enhancement for Higher Accuracy of Loss Differentiation over Small Buffer Heterogeneous Networks". Local Computer Networks, the 34[th] IEEE Conference, Page(s): 249 – 252. 2009

[12] Gong Chang qing,Zhao Linna,Wang Xiaoyan.."Using Neural Network Classifier of Packet Loss Causes to Improve TCP Congestion Control over Ad Hoc Networks". IEEE conference on Microwave, Antenna, Propagation and EMC Technologies for Wireless Communications. , Page(s): 273 – 276 .2007

[13] C. P. Fu and S. C. Liew, "TCP Veno: TCP Enhancement for Transmission over Wireless Access Networks," IEEE J. Sel. Areas Commun., vol. 21,no. 2, pp. 216–228, Feb. 2003.

[14] Eric Hsiao-Kuang Wu, and Mei-Zhen Chen, "JTCP: jitterbased TCP for heterogeneous wireless networks," IEEE Journal on Selected Areas in Communications, vol. 22, no. 4, pp.757-766, May 2004.

[15] Samaraweera N.K.G. (1999) "NonCongestion packet loss detection for TCP error recovery wireless links" Communications, IEE Proceedings-IET JOURNALS

[16] HM El-Sayed, O Bazan, U Qureshi M. Jaseemuddin . (2005) "Performance Evaluation Of Tcp In Mobile Ad-Hoc Networks". The Second International Conference on Innovations in Information Technology

[17] Kim K W, Lorenz P, lee MMO. (2005) "A new tuning maximum congestion window for imporivng TCP performance in MANET". Systems Communications, Proceedings in IEEE Confrences

[18] Jahwan Koo, Sung-Gon Mun, and Hyunseung Choo._(2007) "Sender-Based TCP Scheme for Improving Performance in Wireless Environment". Proceedings of the 7th international conference on Computational Science.